\newcommand{\amindot}[2]{\mbox{#1$\stackrel {\prime}{_{\bf \cdot}}$#2}}
\newcommand{\asecdot}[2]{\mbox{#1$\stackrel {\prime \prime}{_{\bf \cdot}}$#2}}

\documentclass{aa}

\usepackage{graphicx}
\usepackage{txfonts}
\usepackage{longtable}					

\begin{document}

    \title{ISOCAM observations of the \object{L1551} star formation region
	\thanks{Based on observations with ISO, an ESA project with instruments funded by ESA Member
	        States (especially the PI countries: France, Germany, the Netherlands and the United Kingdom)
		and with the participation of ISAS and NASA.}\fnmsep
	\thanks{Based on observations made with the Nordic Optical Telescope, operated on the island of
		La Palma jointly by Denmark, Finland, Iceland, Norway, and Sweden, in the Spanish Observatorio
		del Roque de los Muchachos of the Instituto de Astrofisica de Canarias.}
    }

    \author{M.~G\aa lfalk \inst{1} \and
	   G.~Olofsson \inst{1} \and
	   A.A.~Kaas \inst{2} \and
	   S.~Olofsson \inst{1} \and
	   S.~Bontemps \inst{3} \and
	   L.~Nordh \inst{4} \and
	   A.~Abergel \inst{5} \and
	   P.~Andr\'{e} \inst{6} \and
	   F.~Boulanger \inst{5} \and
	   M.~Burgdorf \inst{13} \and
	   M.M.~Casali \inst{8} \and
	   C.J.~Cesarsky \inst{6} \and
	   J.~Davies \inst{9} \and
	   E.~Falgarone \inst{10} \and
	   T.~Montmerle \inst{6} \and
	   \\ M.~Perault \inst{10} \and
	   P.~Persi \inst{11} \and
	   T.~Prusti \inst{7} \and
	   J.L.~Puget \inst{5} \and
	   F.~Sibille \inst{12}
	   }
    
    \offprints{\\M. G\aa lfalk, \email{magnusg@astro.su.se}}

    \institute{Stockholm Observatory, Sweden				
	  \and Nordic Optical Telescope, Canary Islands, Spain	
	  \and Observatoire de Bordeaux, Floirac, France		
	  \and SNSB, Box 4006, 171 04 Solna, Sweden			
	  \and IAS, Universit\'{e} Paris XI, Orsay, France		
	  \and Service d'Astrophysique, CEA Saclay, Gif-sur-Yvette, France   
	  \and ISO Data Centre, ESA Astrophysics Division, Villafranca del Castillo, Spain   
	  \and Royal Observatory, Blackford Hill, Edinburgh, UK	
	  \and Joint Astronomy Center, Hawaii				
	  \and ENS Radioastronomie, Paris, France				
	  \and IAS, CNR, Rome, Italy						
	  \and Observatoire de Lyon, France					
	  \and SIRTF Science Center, California Institute of Technology, 220-6, Pasadena, CA~91125, USA	
	      }

    \date{Received 28 November 2003 / Accepted 15 March 2004}

    \abstract{

    The results of a deep mid-IR ISOCAM survey of the \object{L1551} dark molecular cloud are presented. The aim of this
    survey is a search for new YSO (Young Stellar Object) candidates, using two broad-band filters centred at
    6.7 and 14.3\,$\mu$m. Although two regions close to the centre of \object{L1551} had to be avoided due to saturation
    problems, 96 sources were detected in total (76 sources at 6.7\,$\mu$m and 44 sources at 14.3\,$\mu$m). Using
    the 24 sources detected in both filters, 14 were found to have intrinsic mid-IR excess at 14.3\,$\mu$m and
    were therefore classified as YSO candidates. Using additional observations in $B$, $V$, $I$, $J$, $H$ and $K$ obtained from
    the ground, most candidates detected at these wavelengths were confirmed to have mid-IR excess at 6.7\,$\mu$m
    as well, and three additional YSO candidates were found.
    Prior to this survey only three YSOs were known in the observed region (avoiding L1551\,IRS\,5/NE and
    \object{HL/XZ\,Tau}). This survey reveals 15 new YSO candidates, although several of these are uncertain due
    to their extended nature either in the mid-IR or in the optical/near-IR observations. Two of the sources
    with mid-IR excess are previously known YSOs, one is a brown dwarf (MHO\,5) and the other is the well known T
    Tauri star \object{HH\,30}, consisting of an outflow and an optically thick disk seen edge on.

    }

    \setlongtables								
    \maketitle

\section{Introduction}

The \object{L1551} dark molecular cloud (Lynds \cite{lynds}) is one of the nearest and therefore most studied regions
of low-mass star formation. It is part of the Taurus-Auriga molecular cloud complex and the distance to its
leading edge has been measured to be 140$\pm$10\,pc (Kenyon et al. \cite{kenyon}).
It shows the usual signs of recent star formation: pre-main-sequence stars (Brice\~{n}o et al. \cite{briceno}), 
Herbig-Haro objects (Devine et al. \cite{devine99}), reflection nebulosity and an extraordinary bipolar outflow 
(Snell et al. \cite{snell}; Rainey et al. \cite{rainey}; Fridlund \& White \cite{fridlund89a}, \cite{fridlund89b}; 
Parker et al. \cite{parker}) that has become a prime example of this outflow type.
This outflow emanates from the IR-source \object{L1551 IRS 5} (Strom et al. \cite{strom}; White et al. \cite{white}; 
Osorio et al. \cite{osorio}), a deeply embedded Young Stellar Object (YSO) found to be a binary (Looney et al. \cite{looney}).
Another well known IR-source with a molecular outflow is \object{L1551 NE}, a binary (Moriarty-Schieven et al. \cite{moriarty}) 
or even triple (Reipurth \cite{reipurth}) YSO, discovered with the IRAS satellite (Emerson et al. \cite{emerson}).
Among the numerous Herbig-Haro objects of the \object{L1551} region, the very compact \object{HH\,30} is a well known YSO with a 
disk and jets that have been imaged by the HST (Burrows et al. \cite{burrows}).
During the ISO (Infrared Space Observatory) mission several nearby dark clouds were surveyed. In this paper we 
present mid-IR observations of the \object{L1551} region using ISOCAM onboard the ISO-satellite.

The low-mass end of the IMF (Initial Mass Function) is a key objective when investigating star formation, 
especially for stars lying in the brown dwarf region (below the Hydrogen burning limit of $0.08\ M_{\sun}$)
where the IMF is not well known. Since \object{L1551} is a nearby star formation cloud, located far away from the 
crowded Galactic plane ($b=-20^{\circ}$), it should be possible to detect such low-mass stars using ISOCAM.
YSO-candidates can generally be found from the ISOCAM data by searching for sources with mid-IR excess, due 
to heated circumstellar dust. These mid-IR observations have high sensitivity and high spatial resolution, 
and are therefore suitable for finding and classifying YSOs, as shown for other clouds 
(Olofsson et al. \cite{olofsson}; Kaas et al. \cite{kaas}; Persi et al. \cite{persi}; Bontemps et al. \cite{bontemps}).

The young stellar population in \object{L1551} has previously been surveyed using several methods: X-ray mapping 
(Carkner et al. \cite{carkner}; Favata et al. \cite{favata}); Optical and near-IR mapping (e.g. Brice\~{n}o et al. \cite{briceno}); 
Optical spectra (Gomez et al. \cite{gomez}) and H$\alpha$ surveys (e.g. Garnavich et al. \cite{garnavich}). 
However, no mid-IR survey as sensitive as this ISOCAM survey (down to $\sim$0.5\,mJy) had previously been done in \object{L1551}.

Lada \& Wilking (\cite{lada}) used observations between 1 and 20\,$\mu$m to plot the mid-IR Spectral Energy
Distribution (SED) of embedded sources in the Ophiuchi dark molecular cloud and found that the population
could be divided into three morphological classes. Later (Lada \cite{lada87}) this YSO classification scheme (Class I--III)
was made more quantitative by introducing the spectral index $\alpha$ and approximate limits for the classes. 
A Class 0 was then introduced (Andr\'{e} et al. \cite{andre93}) and the classification limits were also revised 
(Andr\'{e} \& Montmerle \cite{andre94}).
These four YSO classes are defined in a morphological order with Class 0 sources being protostars at the 
beginning of the main accretion phase, deeply embedded in massive, cold circumstellar envelopes. Class I sources 
are more evolved and Class II YSOs are T Tauri stars with optically thick IR circumstellar disks. 
Class III YSOs represent weak line T Tauri stars with at most optically thin disks, they are therefore hard or 
impossible to detect using mid-IR excess due to their resemblance to field stars. Therefore, using ISOCAM mid-IR 
observations, mainly classes I \& II are detected since Class 0 objects peak in the far-IR spectral region and are 
probably too weak in the mid-IR to be detected by ISOCAM. The evolution from Class 0 to Class III has classically been assumed to
be smooth and gradual, and while this might be the case for isolated young stars it has recently been suggested (Reipurth \cite{reipurth})
that abrupt class transitions can occur in multiple systems due to violent interactions between its members. Massive and quick disk
truncation combined with the possible ejection of light cluster members could then produce highly increased outflow activity as well
as transitions from Class 0 or I objects to visible T Tauri stars.

\section{Observations and data reductions}

\begin{figure*}
	\centering
	\includegraphics[width=17cm]{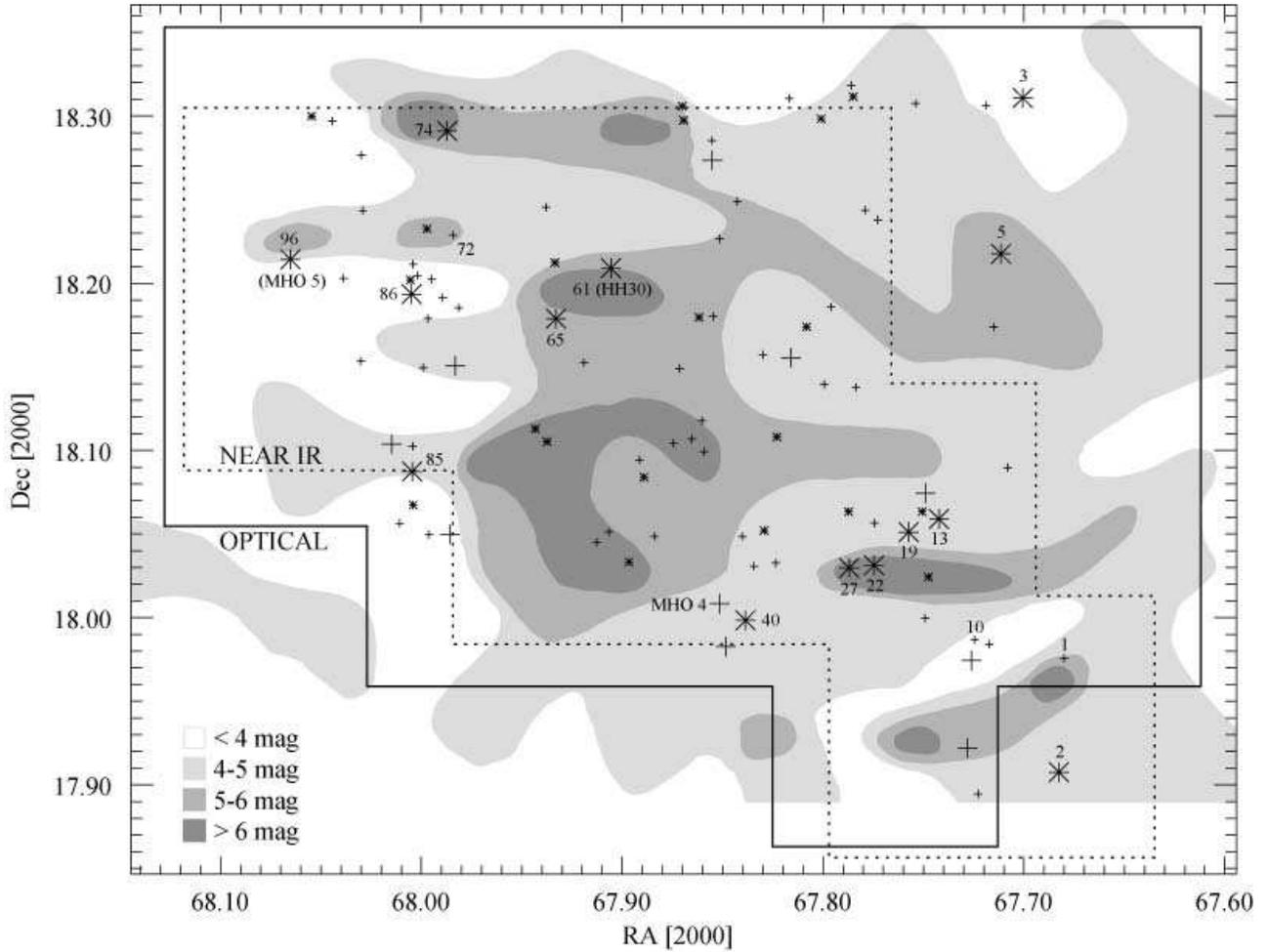}
	\caption{The spatial distribution of all sources detected with ISOCAM, where the large asterisks and 
		plus signs represent `red' and `blue' sources respectively in the $m_{6.7}-m_{14.3}$ colour 
		(see Table~\ref{colour_iso}). The small asterisks and plus signs represent sources only detected 
		at 14.3\,$\mu$m and 6.7\,$\mu$m respectively. The region observed at optical wavelengths 
		($B$, $V$ and $I$) is marked by a solid line while the region observed in the near-IR ($K$) is marked by 
		a dotted line. The contours indicate the visual extinction in magnitude steps (Minn \cite{minn}).
		}
	\label{spatdist}
\end{figure*}

Six regions (total area 0.122\,sq.deg.) were observed 26--27 September 1997 with ISOCAM, avoiding the mid-IR 
bright regions around IRS\,5/NE and HL/XZ Tau due to saturation problems. Two mid-IR filters were used with 
the LW (Long Wavelength) detector, LW2 (6.7\,$\mu$m) and LW3 (14.3\,$\mu$m). For all images, the PFOV is 
6\,$\arcsec$/pixel and the integration time was 2.1\,s for each transmitted frame ($32\times32$ pixels). During the 
reduction, around 19 frames were combined at each frame-position to form images with a temporal history. 
Also, overlaps of half a frame in both equatorial directions gives the much needed spatial redundancy as well 
as additional temporal history. A total of 286 images were used to put together the 6.7 and 14.3\,$\mu$m mosaics. 
Unfortunately the LW detector had a dead column (No\,24), which however could be covered by using overlapping 
images for most parts of the final \object{L1551} mosaic. Only at some parts close to the southern limit of the mosaic 
(due to a spacecraft roll angle of $83^{\circ}$) and close to IRS\,5/NE there is missing information, but 
not more than about 0.3\,\% of the mapped region. Also, the photometry of sources one pixel from the dead 
column and at the mosaic edges is unreliable.

For the data reduction, we used the CIA V4.0 package (Ott et al. \cite{ott}; Delaney et al. \cite{delaney}) 
and the SLICE package (Simple \& Light ISOCAM Calibration Environment) accessed from inside CIA. The 
CIA reduction steps consisted of (in order): Extracting useful observations, dark correction (Vilspa dark 
model, Biviano et al. \cite{biviano}), glitch removal (multiresolution median transform, Starck et al. 
\cite{starck}), short transient correction (Fouks-Schubert model, Coulais \& Abergel \cite{coulais}), 
flatfielding (constant median flatfield from observations).
The data were then processed further using SLICE's long-term transient and variable flatfield algorithms 
(Miville-Deschênes et al. \cite{miville}).
Finally the frames were projected into six raster maps for point source detection and photometry. Each 
detected point source was traced back to its corresponding original frames for temporal and spatial 
verification (in order to exclude remaining artefacts). For the aperture photometry, the point spread 
function was used to correct for flux outside the aperture.
In total 96 sources were detected with photometry possible for 76 sources at 6.7\,$\mu$m and 44 sources at 
14.3\,$\mu$m. Regarding photometric comparison, 24 sources had photometry in both filters and could therefore 
be classified using a colour diagram.
A correlation between the two filters is reasonable to assume for a source separation of less than about one 
pixel ($6\arcsec$), however, to allow for correlations in nebulous regions a limit of \asecdot{8}{5} has been used. 
The mean separation of all correlated sources is however $\sim$ \asecdot{2}{5}.

Even though the integration time was 2.1\,s for each frame and there are about 19 frames for each image, the 
total exposure time for a pixel in the mosaic varies between 40\,s and 160\,s due to the half frame overlaps in 
both RA and Dec which makes the total number of frame-pixels available for a mosaic pixel vary by a factor 
of 1 to 4. Therefore, it is expected that the faintest sources will generally be detected (half a frame, 
16 pixels) away from the mosaic edges. 
From mean uncertainty calculations of all detected ISOCAM sources in \object{L1551}, photometric uncertainties are 
estimated to be ($1\sigma$) 0.4\,mJy at 6.7\,$\mu$m and 0.5\,mJy at 14.3\,$\mu$m which also approximately 
represent the detection limits, however as faint sources as these are only detected in low nebulosity 
regions without artefacts and by using variable flatfielding. On the other hand the surveyed region is not 
completely mapped down to the $1\sigma$ level, since it is possible that even sources brighter than 
$3\sigma$ (1.2\,mJy and 1.5\,mJy respectively) may have been unnoticed due to varying nebulosity, 
glitches, memory effects, uncovered dead columns and source confusion close to very bright sources.
As for the positional accuracy, by using the 19 LW2 sources also seen in the USNO-A2.0 (Monet) catalogue 
the uncertainties have been estimated to 0.13\,s in right ascension and \asecdot{3}{3} in declination.

In order to convert mJy fluxes into 6.7\,$\mu$m and 14.3\,$\mu$m magnitudes, we used the following relations:

\vspace{2mm}

\( \begin{array}{lclcl}
	m_{6.7}  &  =  &  12.30  &  -  &  2.5\log_{10} F_{6.7}  \\
	m_{14.3} &  =  &  10.69  &  -  &  2.5\log_{10} F_{14.3}
   \end{array} \)

\vspace{2mm}

Additional observations were made in H$\alpha$ and the $B$, $V$, $I$ and $K$ bands using the 2.56\,m NOT telescope in La Palma, Canary 
Islands, Spain. The $K$ band observations were obtained 23, 24 and 27 August 1996 using the Arcetri NICMOS3 camera, 
ARNICA, a near-IR (1--2.5\,$\mu$m) $256 \times 256$ pixel HgCdTe array yielding a $2\arcmin \times 2\arcmin$ FOV. The observed 
region covered about the same region as the ISOCAM mid-IR observations and were divided into four mosaics: 
NN~($4\times4$ fields), N~($11\times7$ fields), S~($9\times5$ fields) and SS~($5\times5$ fields). The overlap was \amindot{0}{5} in both 
equatorial directions for all fields. Photometry was done on the individual fields down to the limiting magnitude 
of $K\sim17.5$. There were actually 60 co-added images of 1\,s each in order not to saturate too many sources, giving 
a total exposure time of 60\,s for each field.

\begin{figure}
	\resizebox{\hsize}{!}{\includegraphics{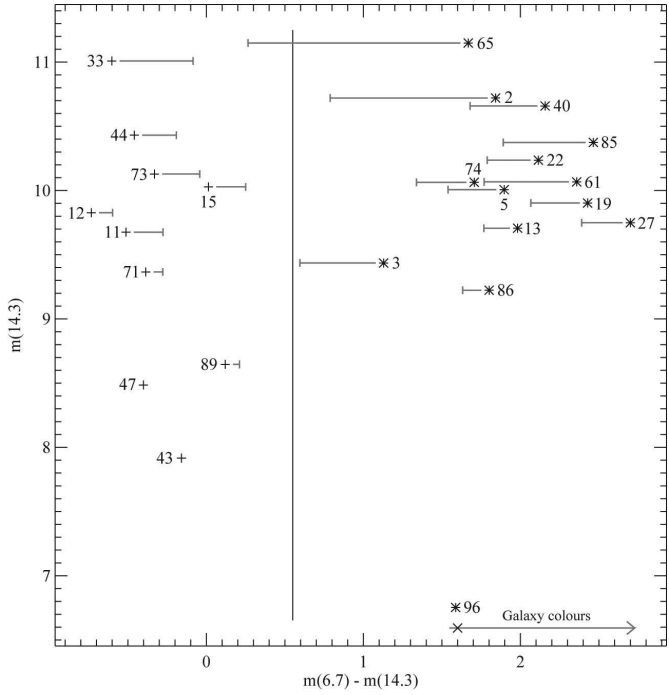}}
	\caption{This mid-IR colour/magnitude plot clearly shows two populations (separated by a dividing line), one
		with a colour index close to zero and one `red' population with 14.3\,$\mu$m excess. The arrow indicates
		expected galaxy colours for ISOCAM observations.
		}
	\label{colour_iso}
\end{figure}

\begin{figure}
	\resizebox{\hsize}{!}{\includegraphics{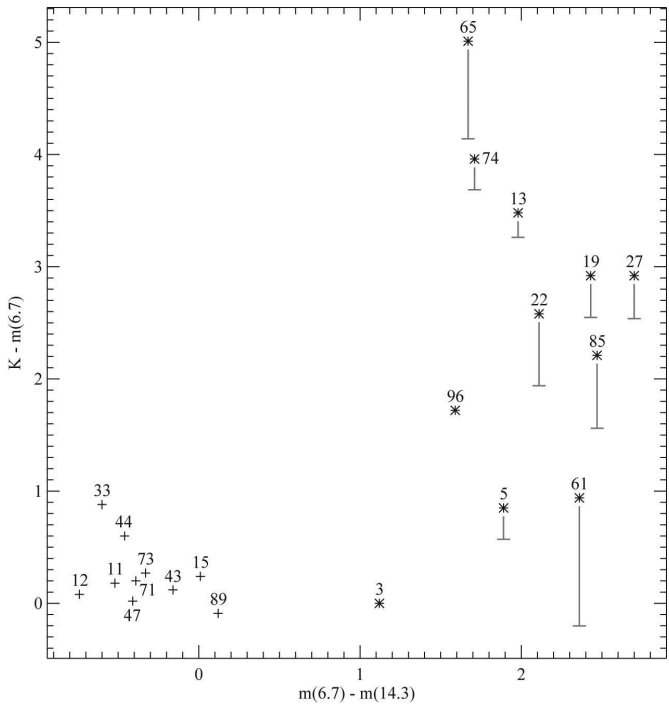}}
	\caption{In this colour/colour diagram for all ISOCAM sources with available $K_S$, 6.7\,$\mu$m and 14.3\,$\mu$m
		magnitudes it is clearly seen that most `red' (asterisks) and `blue' (plus signs) sources
		in the $m_{6.7}-m_{14.3}$ index remain `red' and `blue' separated in the $K-m_{6.7}$ colour index.
		}
	\label{colour_k}
\end{figure}

\begin{figure}
	\resizebox{\hsize}{!}{\includegraphics{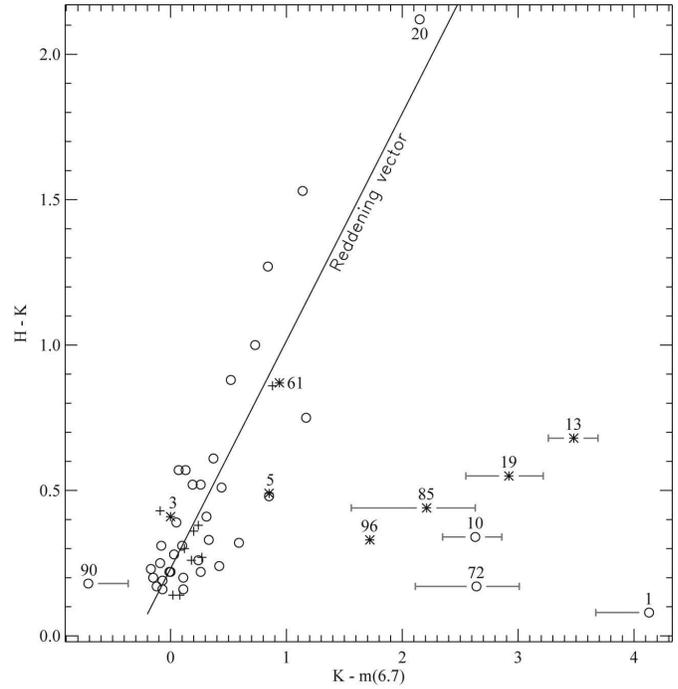}}
	\caption{In this colour/colour diagram for ISOCAM sources with $H$ and $K$ magnitudes, ISO-L1551-1, 10 and
		72 are suggested to be YSO candidates due to excess in the $K-m_{6.7}$ colour index. Symbols are 
		the same as in Fig.~\ref{colour_iso} except that circles represent sources not detected 
		at 14.3\,$\mu$m. The interstellar reddening vector was calculated by fitting a line to the blue
		and previous unclassified sources (circles).
		}
	\label{colour_hk}
\end{figure}

The $B$, $V$, and $I$ observations were carried out on 1 and 3 December 2001 using the ALFOSC instrument (Andalucia Faint 
Object Spectrograph and Camera) mounted on the 2.56\,m NOT telescope. This instrument has a $2048 \times 2048$ pixel CCD chip 
and at a PFOV of \asecdot{0}{188}/pixel it has a FOV of about $\amindot{6}{4} \times \amindot{6}{4}$. We originally set out to do
a $5\times5$ field mosaic of the entire ISOCAM region with an overlap of $45\arcsec$, however, due to bad weather on 3 out 
of the 5 allocated nights we observed the 20 most important frames in the mosaic (regarding ISOCAM source coverage). 
We used the following NOT-filters and exposure times: $B\#74$ (440\,nm, 420\,s), $V\#75$ (530\,nm, 300\,s) and 
$i\#12$ (797\,nm, 180\,s and 10\,s) yielding detection limits of about 22.5, 22.5 and 22.0 mag. respectively. The seeing was 
typically \asecdot{0}{9} for both nights.
Finally, the H$\alpha$ observations were also observed with ALFOSC on the NOT telescope. Two images were obtained on the
night of 22--23 October 2001 using an exposure time of 1200\,s. These were later put together into a mosaic.

\section{Results}

Figure~\ref{spatdist} illustrates the spatial distribution of all sources detected with ISOCAM, where the visual 
extinction in magnitude steps is used to indicate the extent of the \object{L1551} dark molecular cloud (contours 
adopted from Minn \cite{minn}). The distribution appears to be homogeneous, i.e. no apparent correlation between
ISOCAM source positions and visual extinction. The $B$, $V$, $I$ and $K$ observations cover roughly the same region as the ISOCAM 
mosaic, but as the figure shows, one source (ISO-L1551-2) is not included in the optical observations and 15 
sources were missed in the near-IR. However, the 2MASS point source catalogue has been used to obtain near-IR
($J$, $H$ and $K_S$) photometry in the whole ISOCAM region.

\begin{figure*}
	\centering
	\includegraphics[width=18cm]{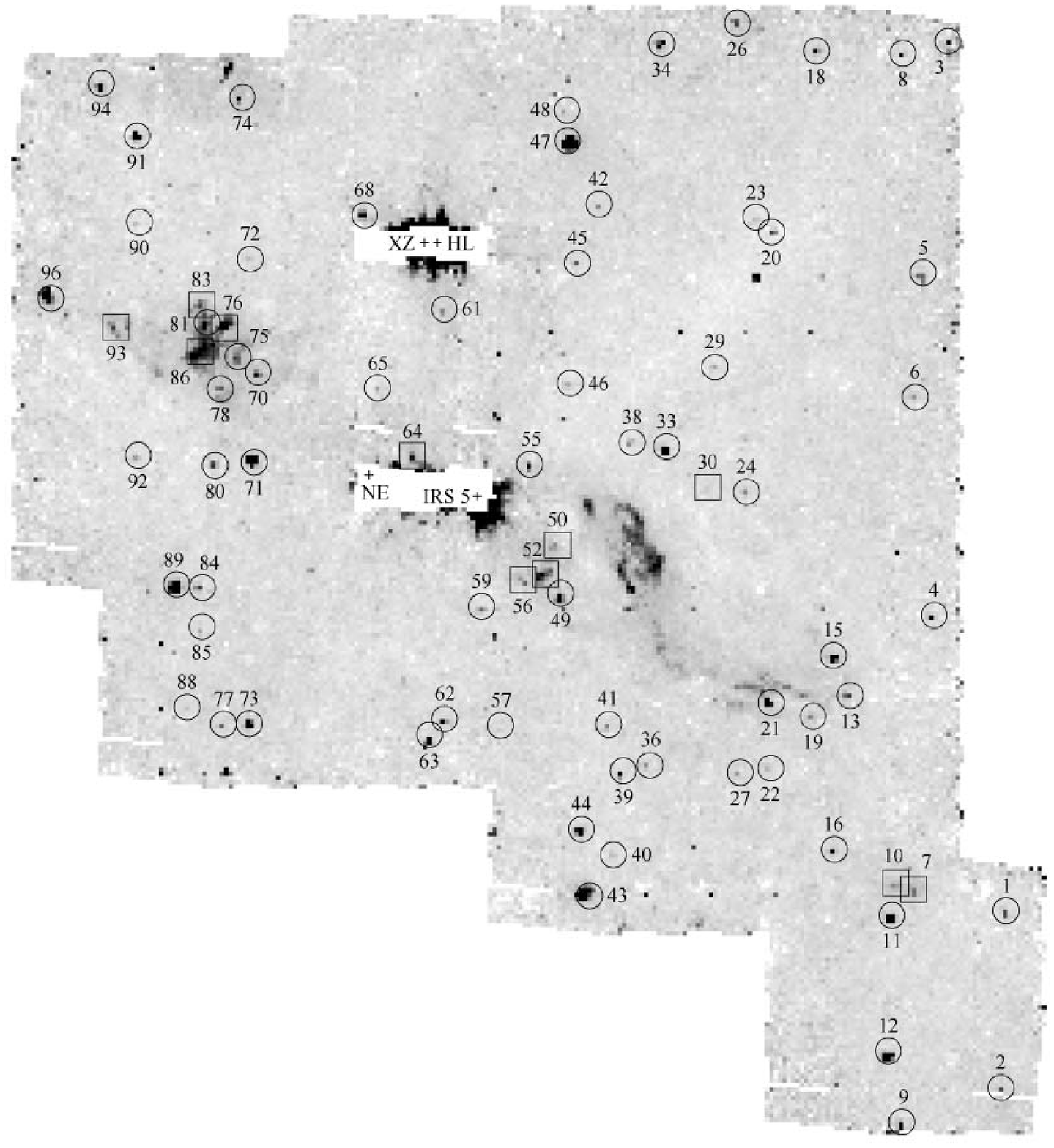}
	\caption{Final 6.7\,$\mu$m ISOCAM map, the image shown is approximately $10\arcmin \times 20\arcmin$.
		Point sources are marked with circles, while squares indicate extended sources (the pixel 
		size is $6\arcsec$). Plus signs mark the positions of \object{L1551 NE}, \object{L1551 IRS5}, HL and \object{XZ Tau}.
	}
	\label{lw2map}
\end{figure*}

The final ISOCAM 6.7\,$\mu$m mosaic is presented in Fig.\,\ref{lw2map} with all 6.7\,$\mu$m sources circled. As 
mentioned previously, two regions (close to IRS5/NE and \object{HL/XZ Tau}) were avoided due to saturation issues.

In Table~\ref{ALLtable}, all detected ISOCAM sources are presented along with optical and near-IR photometry. 
Parenthesis are used to indicate very uncertain sources, where no optical or near-IR counterpart has been 
found and where typically, there is only one overlap available in the 6.7 or 14.3\,$\mu$m mid-IR data. Also, 
bold source numbers indicate sources with an ISOCAM mid-IR excess in the $m_{6.7}-m_{14.3}$ index (i.e. YSO 
candidates). As stated before, the faintest sources should be detected in regions with the most overlaps. 
In fact, almost all sources with a LW2 or LW3 flux below 1\,mJy are located in regions where four 
image-pixels ($4\times19$ frame-pixels) overlap, giving an exposure time of up to 160\,s.

In the ISOCAM mid-IR colour/magnitude plot (Fig.~\ref{colour_iso}), two populations (`red' and `blue') can 
clearly be seen based on the $m_{6.7}-m_{14.3}$ colour index. These populations are well separated by a colour 
index gap larger than one magnitude. Sources with negligible uncertainties in the $m_{6.7}-m_{14.3}$ index, with 
respect to the red-blue dividing line, have no error bars in the plot. Generally, fainter sources at 14.3\,$\mu$m 
have larger colour error bars, with ISO-L1551-65 being the most uncertain `red' source in the plot (it is however 
confirmed as a red source in Fig.~\ref{colour_k}). ISO-L1551-3 has larger error bars than might be expected from 
its 14.3\,$\mu$m magnitude, it has however only one overlap in the ISOCAM mosaics (making its temporal history quite 
uncertain) and is located close to a mosaic edge, making its sky background uncertainty larger as well.

Interstellar reddening in the $m_{6.7}-m_{14.3}$ colour index is small or not even present since `blue' ISOCAM 
(background) sources are also seen in regions with quite large extinction (see Fig.~\ref{spatdist}), therefore 
sources without intrinsic 14.3\,$\mu$m excess should be located close to 0 in the $m_{6.7}-m_{14.3}$ index.
For the stars with near-IR photometry, the intrinsic and interstellar reddening effects can easily be separated 
in a colour/colour diagram using a reddening vector, which indicates the interstellar reddening in the two colours. 
A set of colour/colour diagrams using near-IR and mid-IR photometry are presented in Figures \ref{colour_k} and 
\ref{colour_hk}. The reddening vector in Fig.~\ref{colour_hk} was calculated by fitting a line to the blue and
previously unclassified sources. There is no reddening line in Fig.~\ref{colour_k} since this would 
be an almost vertical line due to the very small interstellar extinction in the $m_{6.7}-m_{14.3}$ colour.
As can be seen from these figures, most sources with red $m_{6.7}-m_{14.3}$ colours are also red in the $K-m_{6.7}$ 
colour and the `blue' sources remain `blue'.
Thus, these ISOCAM mid-IR excesses are confirmed as intrinsic. The $m_{6.7}-m_{14.3}$ red sources ISO-L1551-3 and
61 have no excess at 6.7\,$\mu$m (see Fig.~\ref{colour_hk}). ISO-L1551-61 is however the known YSO HH\,30, consisting of
an outflow and an optically thick disk seen edge on and ISO-L1551-3 is probably a Class\,III YSO and thus has a small
excess (at 14.3\,$\mu$m).
We also note that three sources (ISO-L1551-1, 10 and 72), not detected at 14.3\,$\mu$m, show intrinsic mid-IR
excesses (at 6.7\,$\mu$m). They are therefore added to our list of
sources with mid-IR excesses (i.e. YSO candidates). The ISOCAM `red' source ISO-L1551-2 could not be plotted in the 
colour/colour diagrams since the mid-IR photometry is very uncertain due to its proximity to an uncovered dead 
column in the 6.7\,$\mu$m and 14.3\,$\mu$m mosaics. However, since roughly the same relative flux is lost on the 
dead column for both mid-IR filters, ISO-L1551-2 is still included as a YSO candidate.

\begin{table*}
	\caption{Summary for ISOCAM sources with mid-IR excess (YSO candidates).}
	\label{YSOtable}
	\begin{tabular}{crrcclcl}
	  \hline
        \noalign{\vspace{0.5mm}}
        ISO-L1551 & $F_{6.7}$\,[mJy] & $F_{14.3}$\,[mJy] & $F_{14.3}/F_{6.7}$ & $\alpha_{6.7-14.3}$ & YSO type$^{\mathrm d}$ & Status & Comments \\
        \noalign{\vspace{0.5mm}}
        \hline
	  \multicolumn{8}{c}{\it New YSO candidates} \\
        \hline
	  1 & 1.56 $\pm$ 0.47 & ---~~~~~~~ & --- & $+$0.55$^{\mathrm c}$ & Class II$^{\mathrm c}$ & New & Mid-IR excess at 6.7 $\mu$m (S comp.) \\
        {\bf 2} & 0.79:~~~~~ & 0.97:~~~~ & 1.24: & $-$0.72: & Class II ? & New & Partially on dead column\\
        {\bf 3} & 4.95 $\pm$ 0.68 & 3.18 $\pm$ 1.21 & 0.64 & $-$1.58~ & Class II / III & New & \\
	  {\bf 5} & 1.44 $\pm$ 0.33 & 1.88 $\pm$ 0.43 & 1.30 & $-$0.65~ & Class II & New & \\
	  10 & 2.00 $\pm$ 0.45 & ---~~~~~~~ & --- & $-$0.68$^{\mathrm c}$ & Class II$^{\mathrm c}$ & New & Mid-IR excess at 6.7 $\mu$m \\
        {\bf 13} & 1.76 $\pm$ 0.21 & 2.47 $\pm$ 0.38 & 1.41 & $-$0.55~ & Class II & New & Extended + Close to GH 2$^{\mathrm a}$ \\
 	  {\bf 19} & 0.97 $\pm$ 0.27 & 2.07 $\pm$ 0.43 & 2.12 & $-$0.01~ & Class I / II & New & \\
    	  {\bf 22} & 0.95 $\pm$ 0.29 & 1.52 $\pm$ 0.40 & 1.59 & $-$0.39~ & Class II & New & Extended \\
        {\bf 27} & 0.87 $\pm$ 0.23 & 2.38 $\pm$ 0.37 & 2.73 & $+$0.32~ & Class I & New & Extended \\
	  {\bf 40} & 0.62 $\pm$ 0.23 & 1.03 $\pm$ 0.28 & 1.66 & $-$0.34~ & Class II & New & XMM-Newton-15, Favata et al. \cite{favata}\\
	  {\bf 65} & 0.62 $\pm$ 0.25 & 0.66 $\pm$ 0.47 & 1.05 & $-$0.93~ & Class II & New & VLA 21cm - 15$^{\mathrm b}$ \\
	  72 & 0.87 $\pm$ 0.33 & ---~~~~~~~ & --- & $-$0.68$^{\mathrm c}$ & Class II$^{\mathrm c}$ & New & Mid-IR excess at 6.7 $\mu$m \\
        {\bf 74} & 1.63 $\pm$ 0.33 & 1.78 $\pm$ 0.45 & 1.09 & $-$0.88~ & Class II & New & Extended \\
        {\bf 85} & 0.61 $\pm$ 0.28 & 1.34 $\pm$ 0.43 & 2.20 & $+$0.04~ & Class I / II & New & Double in optical (S comp. extended) \\
        {\bf 86} & 3.24 $\pm$ 0.38 & 3.86 $\pm$ 0.41 & 1.19 & $-$0.77~ & Class II & New & Close to \object{HH 262} \\
        \hline
	  \multicolumn{8}{c}{\it Previously known YSOs} \\
	  \hline
        {\bf 61} & 0.89 $\pm$ 0.58 & 1.77 $\pm$ 0.59 & 1.99 & $-$0.09 & Class I / II & \object{HH 30} & Circumstellar disk seen edge-on \\
        {\bf 96} & 38.34 $\pm$ 0.71 & 37.55 $\pm$ 0.62 & 0.98 & $-$1.03 & Class II & MHO 5 & Spectral type M6 - M6.5 \\
	  \hline

	  \noalign{\medskip}
	  \multicolumn{8}{l}{Note - Bold source numbers indicate YSO candidates based on mid-IR excess emission in the ISOCAM $m_{6.7}-m_{14.3}$ index} \\

	\end{tabular}

	\begin{list}{}{}
		\item[$^{\mathrm{a}}$] Graham \& Heyer \cite{graham}
		\item[$^{\mathrm{b}}$] Giovanardi et al. \cite{giovanardi} - Could be an extragalactic triple radio source 
		   (Rodr\'{\i}guez \& Cant\'{o} \cite{rodriguez83})
		\item[$^{\mathrm{c}}$] Spectral index $\alpha_{2.2-6.7}$ has been used
		\item[$^{\mathrm{d}}$] As implied by $\alpha_{6.7-14.3}$, assuming that all candidates are YSOs
	\end{list}
\end{table*}


In Table~\ref{YSOtable} we list all ISOCAM sources with mid-IR excess and which are therefore YSO candidates. 
Cross correlations with previous source names and YSO status are given. Using the SED indices 
$\alpha_{6.7-14.3}$ and $\alpha_{2.2-6.7}$ all YSO candidates have been classified into YSO 
classes I--III (mostly Class II). For the ISOCAM fluxes, the classical IR spectral index becomes:

\[ \alpha_{6.7-14.3} = \frac{d\log(\lambda F_\lambda)} {d\log\lambda} = 
-\frac{d\log(\nu F_\nu)} {d\log\nu} = \frac{\log(F_{14.3}/F_{6.7})}{\log(14.3/6.7)}-1 \]

and for the $K$ \& 6.7\,$\mu$m fluxes we similarly have:

\[ \alpha_{2.2-6.7} = \frac{\log(F_{6.7}/F_{2.2})} {\log(6.7/2.2)} -1 \]

Classification limits (Class I--III) for both $\alpha$ indices have been assumed to be the same as in many previous ISOCAM studies (e.g. Kaas et 
al. \cite{kaas}; Bontemps et al. \cite{bontemps}). For the ISOCAM spectral index, we thus have: Class I/II limit at $\alpha_{6.7-14.3}\sim0$, 
Class II/III limit at $\alpha_{6.7-14.3}\sim-1.6$ and for no excess (simple photospheric blackbody emission) we have 
$\alpha_{6.7-14.3}\sim-3.0$. For the three YSO candidates without 14.3\,$\mu$m flux, the classification was done using the 
$\alpha_{2.2-6.7}$ index. The total flux uncertainties for the YSO candidates in Table~\ref{YSOtable} were obtained by 
adding the spatial and temporal errors in quadrature. Temporal errors were estimated from frame photometry (in all overlaps) for each source 
position, spatial errors were calculated from the background sky variation close to each source (taking into account the aperture size used). 

Of the 17 YSO candidates, only two (HH\,30 and MHO\,5) were previously known to be YSOs. There is a third previously known YSO (MHO\,4, 
Brice\~{n}o et al. \cite{briceno}) in the ISOCAM region, however, for which we lack evidence of YSO status. It is detected at both 6.7 and 
14.3\,$\mu$m (ISO-L1551-44) but shows no mid-IR excess. This source has also been observed as an X-ray source (\object{L1551 X15}, 
Carkner et al. \cite{carkner}). Some of the new YSO candidates in Table~\ref{YSOtable} are doubtful since they are either extended
or lie close to a known Herbig-Haro object. Also, ISO-L1551-1 has an uncertain YSO status since it was
not detected at 14.3\,$\mu$m and its optical/near-IR counterpart is located more than 8$\arcsec$ from the ISOCAM position. ISO-L1551-2 is
doubtful since it, as described above, lies partially on an uncovered dead column in both ISOCAM mosaics.

\section{Discussion}

\begin{figure*}
	\centering
	\includegraphics[width=17cm]{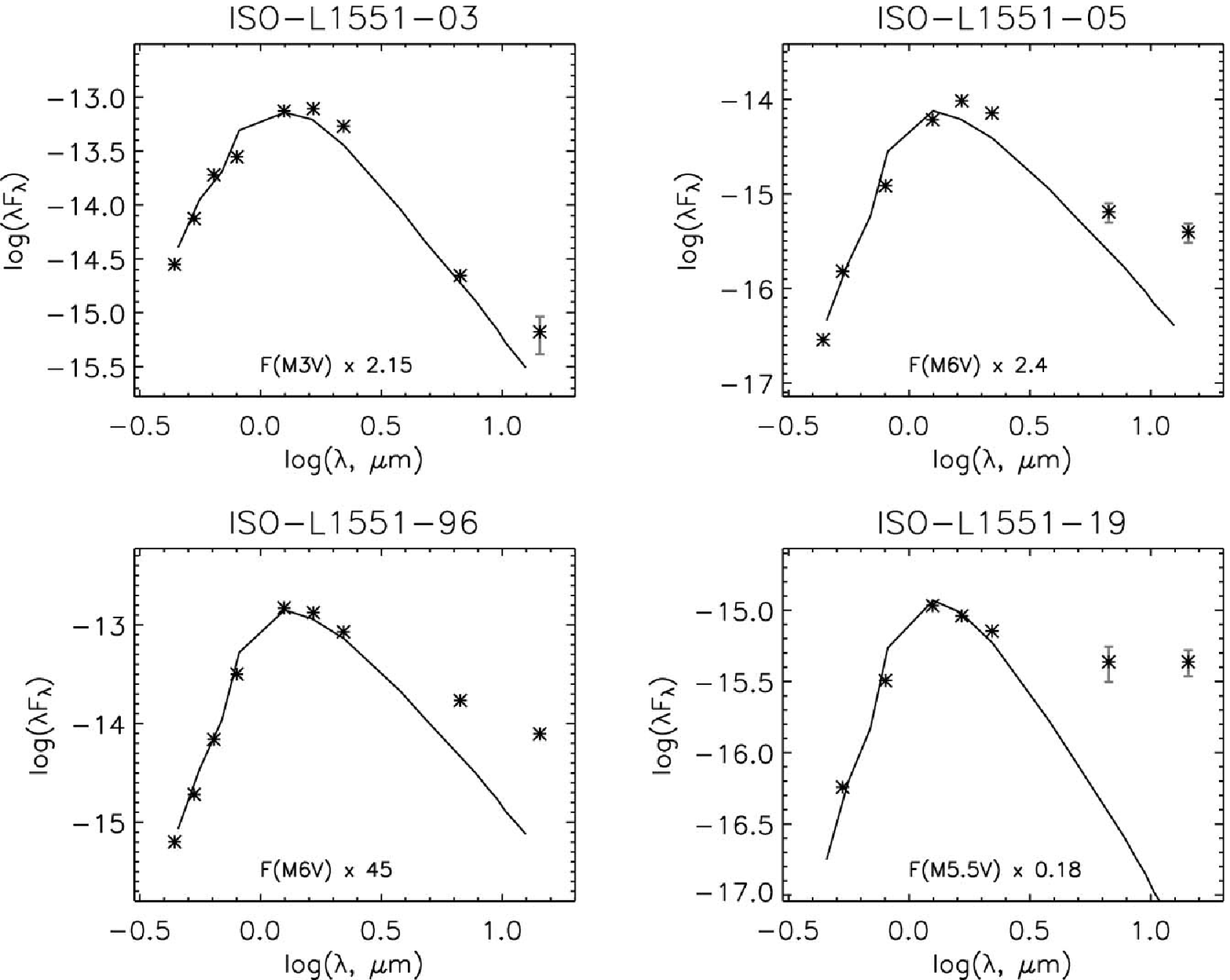}
	\caption{The SEDs of four `red' ISOCAM sources compared to scaled SEDs of M dwarfs.
	In each case an IR excess is clearly seen.
	}
	\label{SED}
\end{figure*}

Although the ISOCAM mid-IR ($m_{6.7}-m_{14.3}$) colour is efficient in separating field stars from YSO candidates, there are mainly 
three other types of objects that could `contaminate' the survey and be mistaken for YSOs. These are: very `red' field stars (AGB 
or M7III spectral types), galaxies and Herbig-Haro objects.

From the Wainscoat et al. (\cite{wainscoat}) model of the point source mid-IR sky, statistically there should be no AGB or M7III 
field stars in our field, these could otherwise be mistaken for YSOs due to their mid-IR brightness.

In Fig.~\ref{colour_iso}, an arrow has been drawn indicating expected $m_{6.7}-m_{14.3}$ colours of galaxies.
For a galaxy with no 14.3\,$\mu$m excess (i.e. $F_{6.7} = F_{14.3}$) 
we expect a colour of $m_{6.7}-m_{14.3} \sim 1.6$ (marked with an x on the arrow). The arrow was adopted from ISOCAM observations 
of 46 non-saturated galaxies (Dale et al. \cite{dale}), including normal star-forming galaxies of many morphological types 
(Ellipticals, Lenticulars, early to late type Spirals, Irregular \& Peculiar types). The bluest of these galaxies are located close 
to the `no 14.3\,$\mu$m excess mark' (Quiescent galaxies) while galaxies with higher star formation rates and `red' morphological 
types are located further along the arrow. There is a probability of about 70\,\% for a randomly observed galaxy to be located within 
the colours of the arrow, while starburst galaxies may be located at much `redder' colours.
From the discussion above, it is clear that differentiation between YSOs and galaxies using only the $m_{6.7}-m_{14.3}$ colour is impossible.
According to the ELAIS (European Large Area ISO Survey) ISOCAM is very sensitive to strongly star-forming galaxies at 14.3\,$\mu$m
(V\"ais\"anen et al. \cite{vaisanen}) and based on ELAIS counts about 6 galaxies are expected in a region of this size at the achieved sensitivity.

Also, many of our ISOCAM sources are only detected at 14.3\,$\mu$m (16 sources; see Table~\ref{ALLtable}) but
since only about 6 galaxies are statistically expected in the region starburst
galaxies can only account for some of these sources, most likely the faint ones due to the steep luminosity function and red colours of such galaxies.
Our source detection limits at 6.7\,$\mu$m and $K$ are about 12.5 and 17.5 mag. respectively.
Yet many of the sources seen only at 14.3\,$\mu$m are very bright (e.g. ISO-L1551-51; $m_{14.3}=7.96$) indicating that these sources are most likely
asteroids. In fact, no starburst galaxies brighter than about $F_{14.3} \sim 5$\,mJy are expected in the whole region and the motion of asteroids would
explain why these bright sources are only detected in one filter. Given the
proximity of \object{L1551} to the ecliptic plane and the large number of asteroids with orbits in this region this seems very likely.

\begin{figure*}
	\centering
	\includegraphics[angle=90,width=18cm]{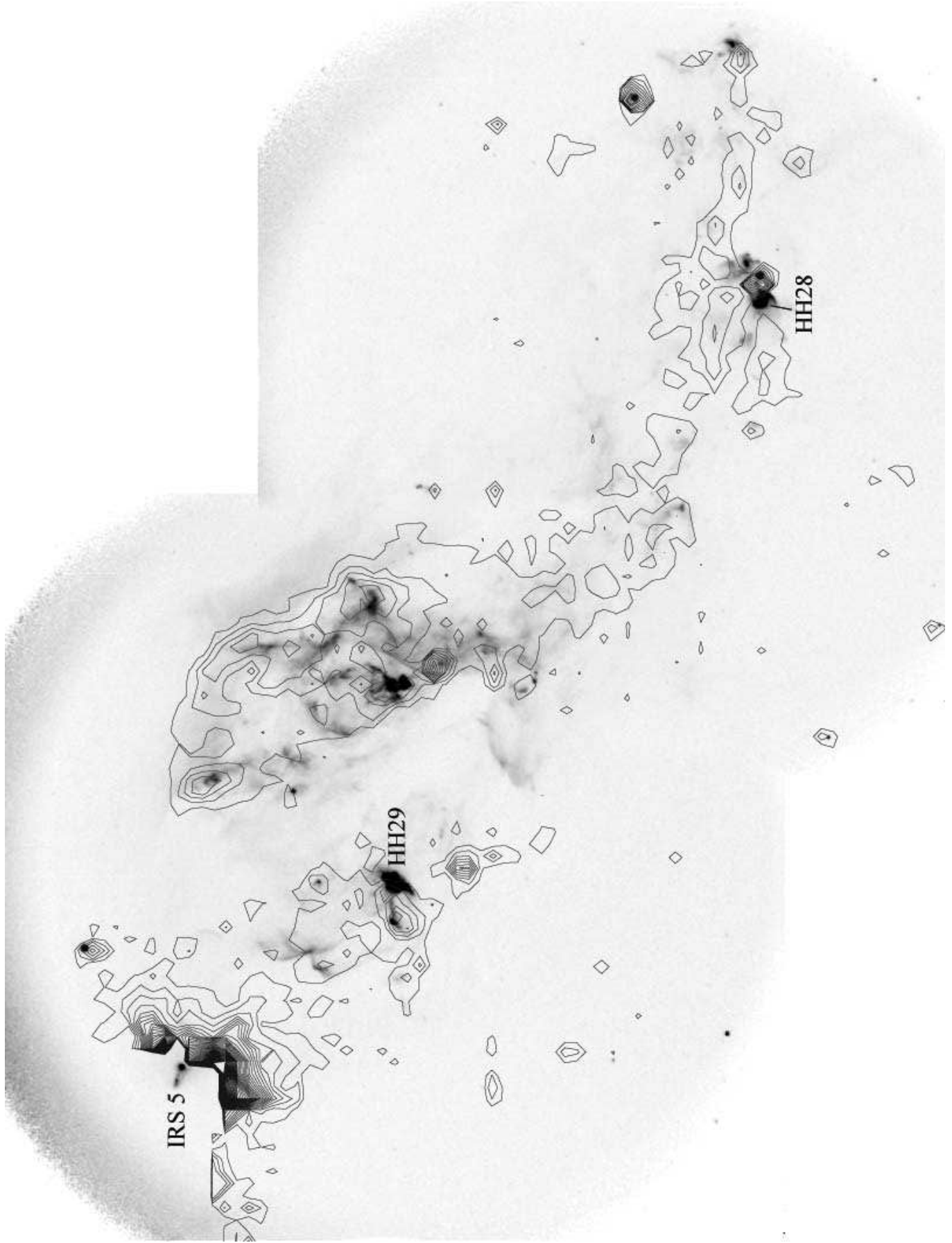}
	\caption{H$\alpha$ mosaic of \object{L1551} observed with the 2.56\,m NOT telescope (exposure time 1200\,s). The contours represent levels of
		constant flux in the ISOCAM 6.7\,$\mu$m map. H$\alpha$ image kindly provided by M.\,Fridlund.
	}
	\label{Halfa}
\end{figure*}

There are numerous known Herbig-Haro objects in the observed region. These shock-excited emission line nebulae are associated with 
outflows from protostellar sources and radiate mainly in H recombination lines and forbidden lines such as 
[S\,II]($\lambda\lambda6717, 6732$), see e.g. HST observations of \object{HH\,29} (Devine et al. \cite{devine00}). At infrared wavelengths we 
could also expect line emission through H$_2$ rovibrational lines and in the ISOCAM filters used for our observations we especially 
have 5 pure rotational H$_2$ lines (H$_2$\,0-0\,S(4) through S(8)) in the 6.7\,$\mu$m filter and 2 pure rotational lines (S(1) and S(2)) 
in the 14.3\,$\mu$m filter. As can be seen in Fig.~\ref{Halfa} the very bright Herbig-Haro objects \object{HH\,28} and \object{HH\,29} 
have 6.7\,$\mu$m peaks that are clearly separated from the corresponding H$\alpha$ peaks. Since the mid-IR and H$\alpha$ traces very
different conditions this could be expected, however, there is also the possibility that extended YSO candidates in fact are galaxies
seen through the dark molecular cloud.
As with galaxies, Herbig-Haro objects cannot be easily separated from YSOs using only the $m_{6.7}-m_{14.3}$ colour. Deep imaging 
or spectra of the new YSO candidates is needed to confirm their nature. Deep images, preferably in the IR (much deeper than our 
$K$ images), could show if the YSO candidates look extended or point source like. Also, one could compare an image taken through an 
[S\,II] interference filter (where HH objects are strong) with a broadband image centred at about 1~micron (where HH objects are 
weak). Spectra of the YSO candidates could be used to search for strong H$\alpha$ emission and Li\,I\,$\lambda6707$ in absorption which 
would indicate a T Tauri star.

\begin{figure*}
	\centering
	\includegraphics[width=17cm]{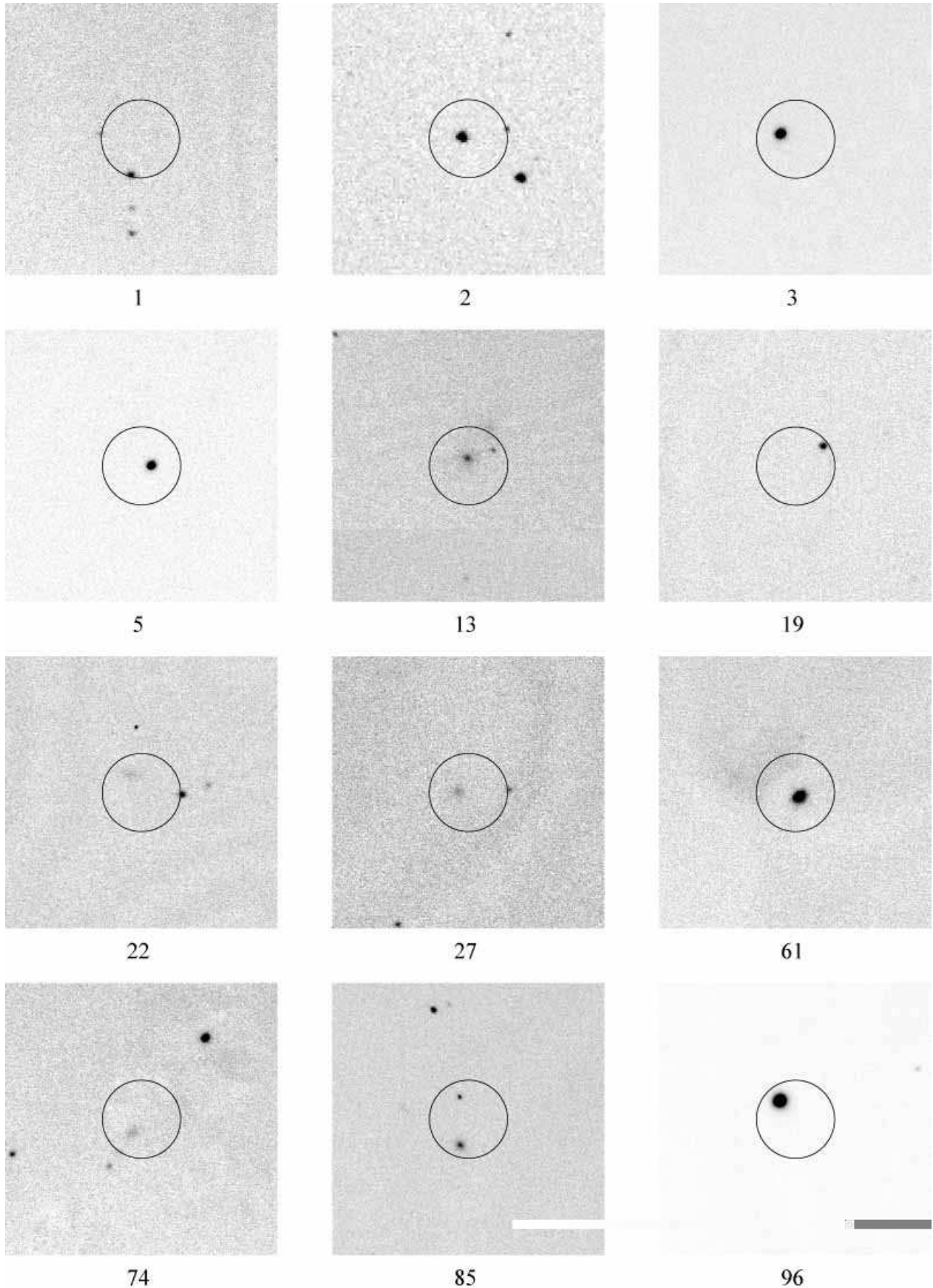}
	\caption{$I$-band images (except for ISO-L1551-2, $K$-band image). All images show a $1\arcmin \times 1\arcmin$ field centred
		on ISOCAM source coordinates where the position uncertainty is indicated with \asecdot{8}{5} radius error circles. ISO-L1551-1 
		and ISO-L1551-85 are both double in the $I$-band images, however, it is probably the south component that has an mid-IR excess
		in both cases.
	}
	\label{Iband_imgs}
\end{figure*}

In Figure~\ref{Iband_imgs}, images are shown for most of the YSO candidates visible in our $I$ or $K$-band observations.
Both ISO-L1551-10 and 72 are easily seen in the optical/near-IR but are not included since they are single point sources well
within the ISOCAM error circle. Several of the YSO candidates are extended in a spherical or elliptical manner, supporting
that these might be galaxies seen through the dark cloud.

At least three of the observed ISOCAM sources have known spectral types from previous studies. ISO-L1551-47 (HD\,285845) is a binary system with
a separation of 73 mas (Schneider et al. \cite{schneider}), excluded from cloud membership by its radial velocity and
proper motion (Walter et al. \cite{walter}). The primary component is of spectral type G8 and has colour indices consistent with that of a main
sequence star (Walter et al. \cite{walter}). Its distance is implied to be 90\,pc from its photometric parallax and it is a bright X-ray
source (Favata et al. \cite{favata}). This suggests that ISO-L1551-47 is an active binary system in the foreground of \object{L1551}.

ISO-L1551-44 (MHO\,4, Brice\~{n}o et al. \cite{briceno}) and ISO-L1551-96 (MHO\,5) are known to be very similar YSOs, both very late type TTS 
of spectral type M6--6.5. They have strong Li\,I\,$\lambda6707$ in absorption and H$\alpha$ in emission and similar ages $\sim1$\,Myr and 
masses $\sim0.05\,M_{\sun}$, placing them in the brown dwarf region. Both were detected at 6.7 and 14.3\,$\mu$m, but only ISO-L1551-96 has 
mid-IR excess at 14.3\,$\mu$m. ISO-L1551-44, however, shows no mid-IR excess at 6.7\,$\mu$m nor in the $K$-band, and is therefore
a very late type NTTS.

ISO-L1551-21 has previously been detected in X-ray observations (L1551\,X13, Carknet et al. \cite{carkner}) where it was suggested to be
unrelated to the cloud since it was identified with \object{LP\,415-1165}, a foreground dM star appearing in the Luyten (\cite{luyten}) proper
motion catalogue. It lies close to the Herbig-Haro objects \object{HH\,28} ($17\arcsec$\,E) and \object{HH\,258} ($20\arcsec$\,NW) but is
definitely a point source.

ISO-L1551-61 (HH\,30) is a very well known YSO, observed as an optically thick circumstellar disk seen edge on. HST observations (Burrows 
et al. \cite{burrows}) have shown clearly that this star is not observed directly, only nebulosity is seen.

ISO-L1551-65 lies at the position of a known triple radio continuum source, probably extragalactic in
nature (Rodr\'{\i}guez \& Cant\'{o} \cite{rodriguez83}) with an overall extent of $\sim20\arcsec$. It is however unclear if
this is the object detected with ISOCAM.

The SEDs for three of the new YSO candidates (ISO-L1551-3, 5, 19) and the known YSO MHO\,5 (ISO-L1551-96) are shown in 
Fig.~\ref{SED}. These sources have at least 7 known broad band magnitudes (see Table~\ref{ALLtable}), making it worthwile 
to plot their SEDs. No extinction corrections have been applied for the new candidates (since no spectral classes are known), 
however, they are all outer members of \object{L1551}, located quite far away from the dense central region so the extinction is probably 
not that large (especially for ISO-L1551-3, see Fig.~\ref{spatdist}). For MHO\,5 the extinction is known to be only $A_V = 0.01$ 
(Brice\~{n}o et al. \cite{briceno}) locating it on our side of the cloud with a negligible extinction. For each source a spectral 
class has been calculated by fitting the observed SEDs with scaled SEDs of M dwarfs as given by Leggett (\cite{leggett}). 
The known brown dwarf MHO\,5 (ISO-L1551-96) is very well fitted using a M6 dwarf, this agrees with the spectral type
M6--6.5 found from spectra of this source (Brice\~{n}o et al. \cite{briceno}). The three YSO candidates have SEDs that mimics dwarfs
of spectral types M3, M6 and M5.5 respectively. All four sources show mid-IR excess when compared to the scaled SEDs.
For ISO-L1551-3 the excess is small and only at 14.3\,$\mu$m which could indicate that the inner part of its accretion disk has been
cleared out while the excess originates further out where the dust is cooler.

\section{Conclusions}

Based on a deep mid-IR ISOCAM survey (approximately $20\arcmin \times 20\arcmin$) of the \object{L1551} dark molecular cloud we have found 14 
sources with intrinsic mid-IR excess emission at 14.3\,$\mu$m which were therefore classified as YSO candidates. Additional 
observations in $B$, $V$, $I$, $J$, $H$ and $K$ supported the YSO candidate status for most detected candidates and yielded three more candidates. 
Out of the 17 detected YSO candidates only two were previously known (\object{HH\,30} and \object{MHO\,5}). This means 15 new
candidates, however, several of these are extended and could be background galaxies or Herbig-Haro objects. 
Assuming a co-eval age of 2-3\,Myr for our sample (mainly Class II objects) and that all candidates actually are YSOs we only add a few solar
masses to the stellar mass component of the Star Formation Efficiency (SFE) in \object{L1551}.
This is however a small contribution when compared to the previously known YSO population.
The new YSO candidates add to the low-mass end of the IMF in \object{L1551}, but due to the small number of known YSOs (as expected
since \object{L1551} has a mass of only $\sim80\,M_{\sun}$, Snell \cite{snell81}) and the lack of follow-up spectra, the IMF can not be
accurately modelled yet.
In addition to the 5 known outflow YSOs (\object{HH\,30}, \object{HL/XZ Tau}, \object{L1551 IRS 5} and \object{L1551\,NE}) two brown dwarfs were 
previously known in the region and one of these, MHO\,5, belongs to the ISOCAM YSO candidates and was found in our study to have mid-IR excesses
compatible with being in a Class II phase of evolution. Most of the YSO candidates, assuming they 
are real, seem to belong to the YSO class II group (Classical T Tauri Stars) and thus have optically thick circumstellar disks at 
mid-IR wavelengths. One of the sources (ISO-L1551-61\,=\,HH\,30) has even been seen as an optically thick edge-on disk with the HST.
Obviously, more follow-up studies from the ground should be made, especially spectroscopic observations of all the new YSO candidates 
which is necessary to confirm their YSO status since some of them could well be background galaxies seen through the dark 
molecular cloud or Herbig Haro objects caused by the known YSOs with outflows, close to the centre of \object{L1551}.

\begin{acknowledgements}
	The Swedish participation in this research is funded by the Swedish National Space Board.
	This publication made use of the NASA/IPAC Infrared Science Archive, which is operated by the Jet Propulsion
	Laboratory, California Institute of Technology, under contract with the National Aeronautics and Space
	Administration, and data products from the Two Micron All Sky Survey, which is a joint project of the University
	of Massachusetts and the Infrared Processing and Analysis Center/California Institute of Technology, funded
	by the National Aeronautics and Space Administration and the National Science Foundation.
	AAK thanks Carlos Baffa, Mauro Sozzi, Ruggero Stanga, and Lenoardo Testi from the Arnica team for the instrument
	support at the NOT in 1996.
\end{acknowledgements}

\onecolumn \onecolumn			
\begin{scriptsize}
\begin{longtable}[l]{cccrrllllllllll}
\multicolumn{15}{l} {\small{\bfseries \tablename\ \thetable{}.}						
	All ISOCAM sources in \object{L1551}, with additional optical \& near-IR photometry.} \\
\label{ALLtable} \\

\hline
\noalign{\vspace{0.5mm}}
\multicolumn{1}{c}{ISO-L1551} & \multicolumn{1}{c}{RA(2000)} & \multicolumn{1}{c}{DEC(2000)} & \multicolumn{1}{r}{$F_{6.7}$}
	& \multicolumn{1}{r}{$F_{14.3}$} & \multicolumn{1}{c}{$B$} & \multicolumn{1}{c}{$V$} & \multicolumn{1}{c}{$R$} & \multicolumn{1}{c}{$I$} 
	& \multicolumn{1}{c}{$J$} & \multicolumn{1}{c}{$H$} & \multicolumn{1}{c}{$K_S$} & \multicolumn{1}{c}{6.7\,$\mu$m}
	& \multicolumn{1}{c}{14.3\,$\mu$m} & \multicolumn{1}{l}{Remarks} \\
\noalign{\vspace{0.5mm}}
\hline
\noalign{\vspace{0.5mm}}
\endfirsthead

\multicolumn{15}{l} {\small{\bfseries \tablename\ \thetable{}.} -- continued from previous page} \\
\noalign{\vspace{2.5mm}}
\hline
\noalign{\vspace{0.5mm}}
\multicolumn{1}{c}{ISO-L1551} & \multicolumn{1}{c}{RA(2000)} & \multicolumn{1}{c}{DEC(2000)} & \multicolumn{1}{r}{$F_{6.7}$}
	& \multicolumn{1}{r}{$F_{14.3}$} & \multicolumn{1}{c}{$B$} & \multicolumn{1}{c}{$V$} & \multicolumn{1}{c}{$R$} & \multicolumn{1}{c}{$I$} 
	& \multicolumn{1}{c}{$J$} & \multicolumn{1}{c}{$H$} & \multicolumn{1}{c}{$K_S$} & \multicolumn{1}{c}{6.7\,$\mu$m}
	& \multicolumn{1}{c}{14.3\,$\mu$m} & \multicolumn{1}{l}{Remarks} \\
\noalign{\vspace{0.5mm}}
\hline
\noalign{\vspace{0.5mm}}
\endhead

\hline \noalign{\vspace{1.5mm}} \multicolumn{15}{l}{{\small Continued on next page...}} \\
\endfoot

\hline
\endlastfoot

	  1 & 04:30:43.1 & 17:58:30 & 1.56 & & & 22.27 & & 19.07 & 16.89 & 16.02 & 15.94 & 11.81 & & Sep \asecdot{8}{2}\,S \\
	    &            &          &      & & &       & & 20.56 &       &       &       &       & & Sep \asecdot{9}{2}\,E \\
	  {\bf 2} & 04:30:43.7 & 17:54:25 & 0.79: & 0.97: & 15.9$^{\mathrm d}$ & out & 14.5 & out & 12.87 & 12.43 & 12.23 & 12.56: & 10.72: & \\
	  {\bf 3} & 04:30:48.4 & 18:18:36 & 4.95 & 3.18 & 17.58 & 16.10 & 14.7 & 13.84 & 11.82 & 10.97 & 10.56 & 10.56 & 09.44 & \\
	  4 & 04:30:50.0 & 18:05:20 & 1.87 & & 20.96 & 18.94 & & 15.99 & 13.38 & 12.32 & 11.75 & 11.62 & & \\
	  {\bf 5} & 04:30:50.9 & 18:13:00 & 1.44 & 1.88 & 22.56 & 20.33 & & 17.24 & 14.54 & 13.24 & 12.75 & 11.90 & 10.01 & \\
	  6 & 04:30:51.7 & 18:10:23 & 0.91 & & & 21.06 & & 17.83 & 15.05 & 13.73 & 13.25 & 12.40 & & \\
	  7 & 04:30:52.1 & 17:59:00 & 2.55 & & 21.19 & 19.83 & & 18.21 & & & 15.92$^{\mathrm e}$ & 11.28 & & \\
	  8 & 04:30:52.8 & 18:18:19 & 1.12 & & & 20.01 & & 16.88 & 14.12 & 12.81 & 12.24 & 12.17 & & \\
	  9 & 04:30:53.3 & 17:53:37 & 3.99 & & 16.61 & 15.22 & 14.0 & 13.36 & 11.97 & 11.28 & 11.06 & 10.80 & & \\
	  10 & 04:30:53.8 & 17:59:09 & 2.00 & & 22.62 & 20.74 & & 17.50 & 15.43 & 14.52 & 14.18 & 11.55 & & near HH\,256 \\
	  11 & 04:30:54.2 & 17:58:24 & 17.96 & 2.55 & 16.72 & sat & 13.1 & 12.40 & 10.55 & 09.60 & 09.34 & 09.16 & 09.68 & GSC1269:485 \\
	  12 & 04:30:54.6 & 17:55:16 & 19.16 & 2.21 & 12.3$^{\mathrm d}$ & sat & 11.0 & 10.47 & 09.59 & 09.31 & 09.17 & 09.09 & 09.83 & GSC1269:261 \\
	  {\bf 13} & 04:30:58.2 & 18:03:29 & 1.76 & 2.47 & & 21.76 & & 18.97 & 16.97 & 15.85 & 15.17 & 11.69 & 09.71 & Galaxy\,? \\
	  14 & 04:30:59.4 & 18:01:24 & & 2.78 & & & & & & & & & 09.58 & \\
	  15 & 04:30:59.8 & 18:04:25 & 8.01 & 1.84 & 17.96 & 16.16 & 14.7 & 13.71 & 11.61 & 10.66 & 10.28 & 10.04 & 10.03 & \\
	  16 & 04:30:59.8 & 17:59:56 & 3.74 & & 17.25 & 15.66 & 14.2 & 13.59 & 12.04 & 11.28 & 10.97 & 10.87 & & \\
	  17 & 04:31:00.2 & 18:03:45 & & 0.99 & 22.59 & 21.41 & & 18.78 & 16.69 & 15.82 & 15.48 & & 10.70 & Sep. \asecdot{8}{9}\,ESE \\
	     &            &          & &      &       & 23.17 & & 20.26 &       &       &       & &       & Sep. \asecdot{3}{6}\,ESE \\
	  18 & 04:31:01.2 & 18:18:23 & 2.00 & & 21.07 & 18.93 & & 16.03 & 13.67 & 12.50 & 11.99 & 11.55 & & \\
	  {\bf 19} & 04:31:01.8 & 18:03:00 & 0.97 & 2.07 & & 21.39 & & 18.69 & 16.42 & 15.80 & 15.25 & 12.33 & 09.90 & \\
	  20 & 04:31:05.7 & 18:14:12 & 1.62 & & & & & & 18.71 & 16.04 & 13.92 & 11.77 & & \\
	  21 & 04:31:05.8 & 18:03:19 & 4.79 & & 17.43 & 15.94 & 15.0 & 13.26 & 11.69 & 11.10 & 10.84 & 10.60 & & L1551\,X13$^{\mathrm a}$ \\
	  {\bf 22} & 04:31:05.9 & 18:01:48 & 0.95 & 1.52 & & & & 19.72 & & & 14.93$^{\mathrm e}$ & 12.35 & 10.24 & Galaxy\,? \\
	  23 & 04:31:07.2 & 18:14:33 & 0.82 & & & & & & 17.68 & 15.19 & 13.66 & 12.52 & & \\
	  24 & 04:31:08.2 & 18:08:12 & 1.18 & & & & & 19.22 & 15.53 & 14.04 & 13.29 & 12.12 & & \\
	  25 & 04:31:08.7 & 18:18:37 & & 4.94 & & & & & & & & & 08.96 & \\
	  26 & 04:31:08.9 & 18:19:02 & 1.59 & & 19.63 & 17.84 & 16.6 & 15.18 & 13.48 & 12.56 & 12.31 & 11.80 & & Double (\asecdot{1}{8}) \\
	  {\bf 27} & 04:31:08.9 & 18:01:43 & 0.87 & 2.38 & & & & 19.93 & & & 15.37$^{\mathrm e}$ & 12.45 & 09.75 & Galaxy\,? \\
	  28 & 04:31:09.0 & 18:03:44 & & 1.51 & & & & & & & 17.7$^{\mathrm e}$ & & 10.24 & \\
	  29 & 04:31:11.2 & 18:11:05 & 0.80 & & & & & & 16.90 & 14.65 & 13.38 & 12.54 & & \\
	  30 & 04:31:11.9 & 18:08:18 & 0.73 & & & & & & & & & 12.64 & & \\
	  31 & 04:31:12.5 & 18:17:49 & & 4.21 & & & & & & & & & 09.13 & \\
	  (32) & 04:31:14.1 & 18:10:22 & & 0.63 & & & & & & & & & 11.20 & \\
	  33 & 04:31:15.9 & 18:09:15 & 5.72 & 0.75 & & 22.61 & & 18.07 & 13.88 & 12.15 & 11.29 & 10.41 & 11.01 & \\
	  34 & 04:31:16.3 & 18:18:34 & 6.10 & & 14.3$^{\mathrm d}$ & sat & 12.5 & 12.37 & 11.24 & 10.65 & 10.45 & 10.34 & & GSC1269:1201 \\
	  35 & 04:31:17.6 & 18:06:24 & & 1.82 & & & & & & & & & 10.04 & HH\,264\,? \\
	  36 & 04:31:17.6 & 18:01:53 & 1.06 & & 22.36 & 20.06 & & 16.78 & 14.08 & 12.94 & 12.42 & 12.23 & & \\
	  37 & 04:31:19.0 & 18:03:02 &      & 0.93 & & & & & & & 16.86$^{\mathrm e}$ & & 10.77 & \\
	  38 & 04:31:19.3 & 18:09:21 & 1.30 & & & & & 20.25 & 15.75 & 13.75 & 12.75 & 12.02 & & Close\,double\,? \\
	  39 & 04:31:20.2 & 18:01:45 & 2.06 & & 19.79 & 18.05 & 16.9 & 15.41 & 13.19 & 12.24 & 11.83 & 11.52 & & \\
	  {\bf 40} & 04:31:21.2 & 17:59:49 & 0.62 & 1.03 & & & & & & & & 12.82 & 10.66 & X-ray source$^{\mathrm f}$\\
	  41 & 04:31:21.6 & 18:02:50 & 1.25 & & 22.00 & 19.92 & & 16.63 & 14.07 & 12.83 & 12.31 & 12.05 & & \\
	  42 & 04:31:22.5 & 18:14:51 & 0.75 & & 18.60 & 17.10 & 15.9 & 15.13 & 13.56 & 12.67 & 12.44 & 12.61 & & \\
	  43 & 04:31:23.5 & 17:58:52 & 65.63 & 12.87 & 14.1$^{\mathrm d}$ & sat & 11.4 & 10.96 & 09.09 & 08.18 & 07.88 & 07.76 & 07.92 & GSC1269:1017\\
	  44 & 04:31:24.3 & 18:00:25 & 8.53 & 1.27 & 20.38 & 18.67 & & 14.50 & 11.65 & 10.92 & 10.57 & 9.97 & 10.43 & MHO\,4$^{\mathrm b}$, X15$^{\mathrm a}$ \\
	  45 & 04:31:24.6 & 18:13:30 & 0.85 & & 18.44 & 16.98 & 15.9 & 15.05 & 13.49 & 12.78 & 12.50 & 12.47 & & Sep. \asecdot{2}{7}\,ENE \\
	     &            &          &      & &       & 21.95 &      & 20.15 &       &       &       &       & & Sep. \asecdot{2}{3}\,WSW \\
	  46 & 04:31:25.3 & 18:10:43 & 0.76 & & & & & & & & & 12.60 & & \\
	  47 & 04:31:25.5 & 18:16:19 & 48.56 & 7.62 & 11.0$^{\mathrm d}$ & sat & 09.9 & sat & 08.67 & 08.24 & 08.10 & 08.08 & 08.49 & HD\,285845 \\
	  48 & 04:31:25.6 & 18:17:02 & 0.44 & & 19.92 & 18.21 & 17.2 & 15.55 & 13.99 & 13.36 & 13.11 & 13.20 & & \\
	  49 & 04:31:26.3 & 18:05:52 & 4.25 & & & 21.98 & & 18.28 & 14.75 & 13.30 & 12.58 & 10.73 & & Sep. \asecdot{3}{3}\,W \\
	     &            &          &      & & 22.39 & 20.60 & & 17.03 & 13.79 & 12.46 & 11.74 & & & Sep. \asecdot{5}{3}\,SE \\
	  50 & 04:31:26.6 & 18:06:59 & 1.40 & & & & & & & & & 11.93 & & HH\,260\,? \\
	  51 & 04:31:27.0 & 18:10:41 & & 12.30 & & & & & & & & & 07.96 & \\
	  52 & 04:31:27.7 & 18:06:19 & 3.86 & & & & & & & & & 10.83 & & HH\,29\,? \\
	  53 & 04:31:29.0 & 18:17:45 & & 4.23 & & & & & & & & & 09.12 & \\
	 (54) & 04:31:29.1 & 18:18:16 & & 1.29 & & & & & & & & & 10.42 & \\
	  55 & 04:31:29.3 & 18:08:51 & 3.11 & & 17.52 & 16.06 & 15.0 & 13.69 & 12.36 & 11.73 & 11.49 & 11.07 & & \\
	 (56) & 04:31:29.9 & 18:06:10 & 1.01 & & & & & & & & & 12.29 & & \\
	 (57) & 04:31:32.1 & 18:02:49 & 0.74 & & & & & & & & & 12.63 & & \\
	  58 & 04:31:33.4 & 18:04:57 & & 1.24 & & & & & & & 16.71$^{\mathrm e}$ & & 10.46 & \\
	  59 & 04:31:34.0 & 18:05:34 & 1.32 & & & 23.24: & & 19.39 & 15.24 & 13.40 & 12.52 & 12.00 & & \\
	 (60) & 04:31:35.2 & 18:01:54 & & 0.58 & & & & & & & & & 11.27 & \\
	  {\bf 61} & 04:31:37.5 & 18:12:26 & 0.89 & 1.77 & 19.78 & 18.88 & 16.4 & 17.06 & 15.18 & 14.24 & 13.37 & 12.43 & 10.07 & HH\,30 \\
	  62 & 04:31:37.6 & 18:02:58 & 1.67 & & 19.90 & 18.14 & 17.2 & 15.72 & 13.73 & 12.66 & 12.34 & 11.75 & & \\
	  63 & 04:31:39.0 & 18:02:36 & 2.26 & & 17.33 & 15.99 & 15.1 & 14.03 & 12.32 & 11.65 & 11.34 & 11.42 & & \\
	 (64) & 04:31:40.7 & 18:09:03 & 2.36 & & & & & & & & & 11.37 & & \\
	  {\bf 65} & 04:31:44.1 & 18:10:37 & 0.62 & 0.66 & & & & & & & 17.8$^{\mathrm e}$ & 12.82 & 11.15 & VLA 21cm-15$^{\mathrm c}$$^{\mathrm g}$ \\
	 (66) & 04:31:44.2 & 18:12:38 & & 1.07 & & & & & & & & & 10.62 & \\
	  67 & 04:31:45.0 & 18:06:12 & & 2.82 & & & & & & & & & 09.57 & \\
	  68 & 04:31:45.3 & 18:14:37 & 11.08 & & 17.55 & 16.28 & 15.1 & 14.13 & 11.86 & 10.67 & 10.06 & 09.69 & & \\
	  69 & 04:31:46.5 & 18:06:40 & & 0.84 & & & & & & & & & 10.88 & \\
	  70 & 04:31:55.7 & 18:11:00 & 1.60 & & 20.18 & 18.33 & 17.0 & 15.27 & 13.31 & 12.45 & 12.12 & 11.79 & & \\
	  71 & 04:31:56.1 & 18:08:55 & 21.28 & 3.39 & 17.08 & 15.13 & 13.9 & 12.70 & 10.63 & 09.54 & 09.18 & 08.98 & 09.37 & \\
	  72 & 04:31:56.4 & 18:13:36 & 0.87 & & 20.82 & 19.48 & 19.1 & 17.60 & 16.21 & 15.26 & 15.09 & 12.45 & & \\
	  73 & 04:31:56.6 & 18:02:52 & 10.02 & 1.68 & 16.42 & 15.11 & 13.6 & 12.96 & 11.23 & 10.34 & 10.07 & 09.80 & 10.13 & GSC1269:559 \\
	  {\bf 74} & 04:31:57.2 & 18:17:21 & 1.63 & 1.78 & & 20.50 & & 18.69 & & & 15.73$^{\mathrm e}$ & 11.77 & 10.06 & Galaxy\,? \\
	  75 & 04:31:57.6 & 18:11:21 & 1.81 & & & & & & & & & 11.66 & & \\
	  76 & 04:31:58.9 & 18:12:01 & 4.29 & & & & & & & & 14.97$^{\mathrm e}$ & 10.72 & & \\
	  77 & 04:31:59.1 & 18:02:52 & 0.85 & & 17.43 & 16.46 & 15.7 & 14.70 & 13.31 & 12.69 & 12.47 & 12.48 & & \\
	  78 & 04:31:59.4 & 18:10:37 & 0.60 & & 17.00 & 15.91 & 15.4 & 14.51 & 13.36 & 12.90 & 12.73 & 12.85 & & \\
	  79 & 04:31:59.5 & 18:13:49 & & 6.67 & 21.33 & 20.10 & 19.2 & 18.47 & & & & & 08.63 & Sep. 11$\arcsec$\,N \\
	     &            &          & &      &       & 22.87 &      & 19.90 & & & & &      & Sep. 14$\arcsec$\,NNW \\
	  80 & 04:31:59.9 & 18:08:50 & 2.00 & & 15.51 & sat & 13.9 & 13.25 & 12.13 & 11.67 & 11.48 & 11.55 & & GSC1269:777 \\
	  81 & 04:32:00.6 & 18:12:09 & 2.69 & & & & & & & & & 11.23 & & \\
	  82 & 04:32:01.0 & 18:03:55 & & 1.48 & & & & & & & & & 10.26 & \\
	  83 & 04:32:01.1 & 18:12:33 & 2.42 & & & & & & & & & 11.34 & & HH\,/\,Galaxy ? \\
	  84 & 04:32:01.1 & 18:06:01 & 1.25 & & 20.25 & 18.42 & 17.4 & 15.86 & 13.63 & 12.50 & 12.11 & 12.06 & & \\
	  {\bf 85} & 04:32:01.1 & 18:05:07 & 0.61 & 1.34 & 22.39 & 20.91 & & 18.88 & 16.63 & 15.49 & 15.05 & 12.84 & 10.37 & Sep. \asecdot{6}{0} SSE$^{\mathrm h}$ \\
	     & & & & & & 22.91 & & 19.86 & & & & & & Sep. \asecdot{5}{3} NNE \\
	  {\bf 86} & 04:32:01.3 & 18:11:29 & 3.24 & 3.86 & & & & & & & & 11.02 & 09.22 & HH 262 ? \\
	  87 & 04:32:01.5 & 18:11:59 & & 2.22 & & & & & & & & & 09.83 & HH\,object ? \\
	  88 & 04:32:02.6 & 18:03:15 & 0.55 & & 17.44 & 16.18 & 15.2 & 14.72 & 13.52 & 12.99 & 12.79 & 12.94 & & \\
	  89 & 04:32:03.6 & 18:06:06 & 25.93 & 6.57 & 17.15 & 15.14 & 13.5 & 12.51 & 10.26 & 09.11 & 08.68 & 08.77 & 08.65 & GSC1270:357 \\
	  90 & 04:32:07.2 & 18:14:28 & 0.41 & & 17.70 & 16.39 & 15.4 & 14.74 & 13.45 & 12.75 & 12.57 & 13.28 & & \\
	  91 & 04:32:07.4 & 18:16:28 & 4.54 & & 16.24 & sat & 13.7 & 13.05 & 11.64 & 10.84 & 10.62 & 10.66 & & GSC1270:1234 \\
	  92 & 04:32:07.4 & 18:09:05 & 0.82 & & 16.54 & 15.55 & 14.7 & 14.20 & 13.07 & 12.60 & 12.44 & 12.51 & & \\
	  93 & 04:32:09.5 & 18:12:02 & 1.67 & & & & & & & & & 11.74 & & HH\,/\,Galaxy ? \\
	  94 & 04:32:10.9 & 18:17:41 & 3.26 & & 14.0$^{\mathrm d}$ & sat & 12.8 & 12.54 & 11.63 & 11.29 & 11.13 & 11.02 & & GSC1270:401 \\
	 (95) & 04:32:13.4 & 18:17:51 & & 0.80 & & & & & & & & & 10.93 & \\
	  {\bf 96} & 04:32:15.8 & 18:12:43 & 38.34 & 37.55 & 19.20 & 17.58 & 15.8 & 13.70 & 11.07 & 10.39 & 10.06 & 08.34 & 06.75 & MHO\,5$^{\mathrm b}$ \\

\end{longtable}

\begin{list}{}{}
	\item[$^{\mathrm{a}}$] Carkner et al. \cite{carkner}
	\item[$^{\mathrm{b}}$] Brice\~{n}o et al. \cite{briceno}
	\item[$^{\mathrm{c}}$] Giovanardi et al. \cite{giovanardi} -- Could be an extragalactic triple radio source 
		(Rodr\'{\i}guez \& Cant\'{o} \cite{rodriguez83})
	\item[$^{\mathrm{d}}$] USNO A-2 catalogue has been used
	\item[$^{\mathrm{e}}$] K-band magnitude (ARNICA/NOT)
	\item[$^{\mathrm{f}}$] XMM-Newton-15, Favata et al. \cite{favata}
	\item[$^{\mathrm{g}}$] XMM-Newton-35
	\item[$^{\mathrm{h}}$] XMM-Newton-49
\end{list}

\end{scriptsize}


\end{document}